\documentclass[aps,prl,groupedaddress,twocolumn]{revtex4}

\usepackage{graphicx}
\usepackage{amsmath}
\usepackage{amsfonts}
\usepackage{amssymb}

\usepackage{subfigure}

\usepackage{color}

\begin{document}


\title{Proposed robust and high-fidelity preparation of excitons and biexcitons
        in semiconductor quantum dots making active use of phonons}

\author{M. Gl\"assl$^{1}$}
\email[]{martin.glaessl@uni-bayreuth.de}
\author{A. M. Barth$^{1}$}
\author{V. M. Axt$^{1}$}

\affiliation{$^{1}$Institut f\"{u}r Theoretische Physik III, Universit\"{a}t Bayreuth, 95440 Bayreuth, Germany}

\date{\today}

\begin{abstract}
It is demonstrated how the exciton and the biexciton state of a quantum dot 
can be prepared with high fidelity on a picosecond time-scale 
by driving the dot with a strong laser pulse that is tuned above the exciton
resonance for exciton preparation and in resonance with the exciton transition
for biexciton preparation. The proposed schemes make use of the phonon-induced
relaxation towards photon dressed states in optically driven quantum dots and
combine the simplicity of Rabi flopping with the robustness of adiabatic rapid
passage schemes. Our protocols allow for an on-demand, fast and almost perfect 
state preparation even at strong carrier-phonon interaction where other schemes
fail. In fact, the performance of the presented protocols is shown to be the better
the stronger the carrier-phonon interaction is.
\end{abstract}

\maketitle


Realizing a high-quality, robust, on-demand,  and fast exciton or
biexciton preparation in semiconductor quantum dots (QDs)
is of great importance for many promising QD-based devices such as
single \cite{michler:00,press:07} or entangled
\cite{moreau:01,stevenson:06,akopian:06,dousse:10} photon sources that
are crucial for various applications in the field of quantum information
processing \cite{bouwmeester:00,nielsen:00} as well as for tests of fundamental
aspects of quantum mechanics \cite{haroche:06}. While a perfect initiation
of the QD in both states can in principle be realized via Rabi flopping 
\cite{zrenner:02,stufler:06}, these simple schemes suffer
from a high sensitivity on the dipole moments and the pulse 
intensity. A preparation that is robust against fluctuations
in the coupling strengths can be achieved using protocols with frequency swept
pulses that rely on adiabatic rapid passage (ARP) \cite{schmidgall:10,simon:11,wu:11}.
However, the degree of exciton inversion realized in ARP-based experiments
\cite{simon:11,wu:11} stayed considerably below the ideal case and most recent
theoretical works gave compelling evidence that this reduction can be
attributed to acoustic phonon coupling \cite{lueker:12,debnath:12}
that is also known to strongly limit the fidelity of Rabi flopping  
\cite{foerstner:03,vagov:07,ramsay:10}.

In this Letter, we shall present protocols that combine the simplicity of
Rabi flopping with the robustness of ARP-schemes and give
the discussion of phonon influences a completely different perspective by
demonstrating how one can highly benefit from the otherwise undesired
carrier-phonon coupling. To be specific, we propose protocols allowing for
a high-fidelity and robust phonon-assisted preparation of the exciton as
well as the biexciton state in strongly confined QDs that rely on exciting
the system with an off-resonant intense pulse. Our schemes make active use
of the acoustic phonon coupling by exploiting the characteristics of the
stationary non-equilibrium state towards which the QD is driven due to the
system-environment interaction and are shown to perform the better the
stronger the carrier phonon coupling is.


Let us first concentrate on preparing the single exciton state. To this
end, we assume a circularly polarized excitation, which allows us to
model the laser-driven QD as an electronic two-level system that
consists of the ground state $|G\rangle$ without electron-hole pairs
and the single exciton state $|X\rangle$ described by the Hamiltonian
\begin{align}
H_{\rm{QD,\,L}} \!=\! \hbar\omega_X |X\rangle\langle X| 
          \!-\!\left[\frac{\hbar f(t)}{2} e^{-i\omega_L t} |X\rangle\!\langle G|+\rm{h.c.}\right]\!.
\label{H1}
\end{align}
$\omega_L$ is the laser frequency and $f(t)$ denotes the instantaneous
Rabi-frequency that for a Gaussian pulse is given by
$f(t)\!=\alpha/(\sqrt{2\pi}\tau_0)\exp(-t^2/2\tau_0^2)$, where $\alpha$ is the
pulse area and $\tau_0$ defines the pulse length. The pure dephasing
carrier-phonon coupling is modeled by the Hamiltonian
\begin{align}
H_{\rm{QD, \, Ph}} \!= \!\! \sum_{\bf q} \hbar\omega_{\bf q}\,b^\dag_{\bf q} b_{\bf q} 
\!+\! \sum_{{\bf q}, \nu} \hbar n_{\nu} \big(g_{\bf q} b_{\bf q} \!+\! g^{*}_{\bf q} b^\dag_{\bf q} \big)
|\nu\rangle\langle\nu|,
\label{H2}
\end{align}
where $\nu$ labels the electronic states, $n_{\nu}$ counts the excitons
present in the state $|\nu\rangle$, $b^\dag_{\bf q}$ creates a longitudinal
acoustic (LA) bulk phonon with wave vector ${\bf q}$ and energy
$\hbar\omega_{\bf q}$ and $g_{\bf q}$ denote the exciton-phonon coupling
constants. We concentrate on the deformation potential coupling to LA
phonons that is dominant for GaAs QDs \cite{krummheuer:02,ramsay:10}, we
choose a radial electron confinement length of 3 nm and use the same
coupling parameters as in Ref.~\onlinecite{vagov:04} that have been shown
to nicely reproduce experimental results \cite{vagov:04}. To calculate the
coupled carrier-phonon dynamics, we shall apply a numerically exact
real-time path-integral approach that allows us to study the system
evolution without invoking any approximations to the model given above.
Details of this method can be found in Ref.~\onlinecite{vagov:11}.


The phonon coupling strongly affects the QD dynamics. Most important for
our present discussion, it pushes the laser driven electronic system towards
a stationary non-equilibrium state. We demonstrated in
Ref.~\onlinecite{glaessl:11b} that for a two-level system with weak carrier-phonon
coupling as realized in GaAs QDs and a constant optical driving with $f(t)=\rm{const.}$, 
this stationary state can be well approximated by a thermal distribution over
the eigenstates of $H_{\rm{QD,\,L}}$, often referred to as photon dressed states,
leading to a stationary exciton occupation of
\begin{align}
\label{dressed}
 C_X(t=\infty) = \frac{1}{2}\left[ 1+
                 \frac{\Delta}{\hbar\Omega} \,\tanh{\left( \frac{\hbar\Omega}{2k_BT} \right)} \right] ,
\end{align}
\begin{figure}[ttt]
 \includegraphics[width=8.5cm]{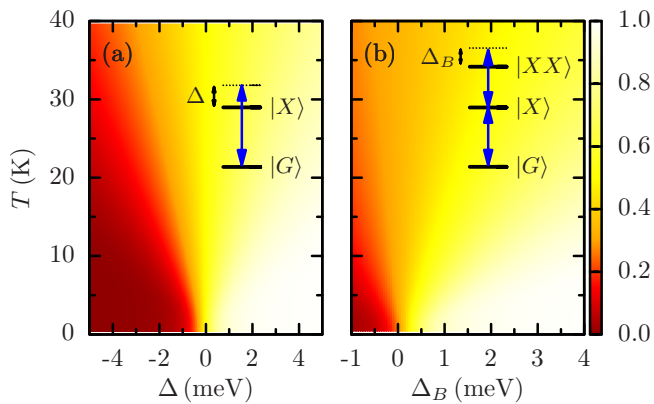}
    \caption{
             (color online) Stationary (a) exciton and (b) biexciton occupation that is reached
             under constant optical driving with $f=1.0\,\rm{ps}^{-1}$ as a function of
             (a) the temperature and the detuning $\Delta$ between the laser frequency
             and the QD transition frequency and (b) the temperature and the biexciton
             binding energy $\Delta_B$. Inset: Sketch of the excitation schemes (see text).
            }
 \label{fig1}
\end{figure}
where $\Omega=\sqrt{f^2+(\Delta/\hbar)^2}$ and $\Delta=\hbar(\omega_L-\omega_X)$
is the detuning between the laser and the QD transition. Obviously,
Eq.~\eqref{dressed} predicts a stationary inversion for
$\Delta/(\hbar\Omega) \approx 1$ and $\hbar \Omega / 2kT \gg 1$. Indeed, the exact
stationary exciton occupation as calculated by using the path integral approach
and shown in Fig.~\ref{fig1}(a) as a function of $\Delta$ and $T$ is almost
perfectly described by Eq.~\eqref{dressed}: while for red-shifted excitation with
$\Delta < 0$ the stationary occupation is below $1/2$, an almost perfect stationary
inversion is found at low $T$ for a blue-shifted laser, similar as previously
predicted for a two-level system consisting of two tunnelling-coupled QDs within
a Born-Markov approximation \cite{stace:05}. However, for the system considered
here, several comments are in order: First, the time needed to reach the
stationary state strongly depends on the excitation conditions and can be up
to several hundreds of picoseconds at small $f$, low $T$ and large $\Delta$.
On these time-scales other relaxation processes like radiative decay, that are
not included in our model, gain in importance and reduce the accessible degree
of inversion \cite{borri:01}. Further, cw-excitations as assumed in Fig.~\ref{fig1}
and Eq.~\eqref{dressed} are of no use when an on-demand inversion is the target.
Instead, pulsed excitations are needed to complete a preparation at a given time.
Therefore, to evaluate whether the stationary inversion as shown in Fig.~\ref{fig1}(a)
has relevance for state preparation, several questions have to be answered, e.g.:
which degree of inversion can be reached for off-resonant pulses of finite length?
What determines the optimal detuning  and how sensitive is the achieved inversion
against variations in the pulse intensity? In this Letter, we will demonstrate
that by applying an off-resonant intense pulse at low temperatures it is not only
possible to achieve some inversion (as achieved in recent experiments with
moderate pulse intensities \cite{ramsay:11}) but an almost perfect state preparation
on a time-scale of several picoseconds.

\begin{figure}[ttt]
 \includegraphics[width=8.3cm]{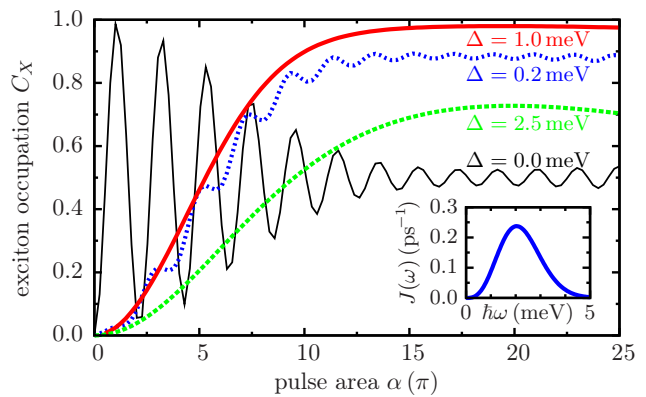}
    \caption{
             (color online) Final exciton occupation $C_X$ after a Gaussian pulse of 15 ps FWHM 
             at $T=4$~K as a function of the applied pulse area
             $\alpha$ for different detunings $\Delta$ as indicated.
             Inset: Phonon spectral density for a spherical GaAS QD with a
             radial electron confinement length of 3 nm. 
            }
 \label{fig2}
\end{figure}

\begin{figure}[ttt]
 \includegraphics[width=8.3cm]{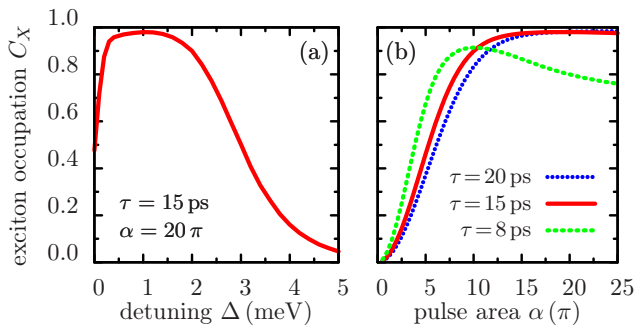}
    \caption{
             (color online) Final exciton occupation $C_X$ after a Gaussian pulse with pulse
             length $\tau$ (FWHM) at $T=4\,\rm{K}$. (a) $C_X$ as a function of $\Delta$ for
             $\tau=15\,\rm{ps}$ and $\alpha=20\pi$. (b) $C_X$ as a function of
             $\alpha$ for $\Delta=1\,\rm{meV}$ and different $\tau$ as indicated.  
            }
 \label{fig3}
\end{figure}

Shown in Fig.~\ref{fig2} is the exciton occupation $C_X$ after a Gaussian pulse of
$\tau = 2\sqrt{2\ln{2}}\,\tau_0 = 15$ ps full width at half maximum (FWHM) as a
function of the pulse area $\alpha$ for different detunings $\Delta$ at $T=4\,\rm{K}$.
While for $\Delta=0$ the occupation performs damped Rabi-oscillations around a mean
value of $1/2$, the results do already considerably change for a slightly
blue-shifted excitation with $\Delta=0.2\,\rm{meV}$: not only does the oscillation
amplitude decrease (as it is well-known for off-resonant driving \cite{allen:75}), but
at high pulse areas, where the phonon damping is stronger \cite{foerstner:03,vagov:07}
and the system is efficiently pushed towards its stationary state, significantly higher 
occupations are reached with $C_X$ taking values of roughly $0.9$. Even higher
occupations are realized for larger $\Delta$. For $\Delta=1.0\,\rm{meV}$, which turns out
to be the optimal choice for the detuning, an almost ideal and robust exciton preparation
is realized at high pulse areas. We would like to stress that an almost equally 
high inversion is found for a rather wide range of detunings as shown in Fig.~\ref{fig3}(a),
where $C_X$ is plotted as a function of $\Delta$ for a fixed pulse area of $\alpha=20\pi$.
Thus, the  phonon-assisted state preparation as realized by exciting the system off-resonantly
by a single intense pulse is not only of high fidelity but also robust against variations
in the pulse intensity or the detuning provided that the pulse is strong enough.

It should be noted, that although the stationary occupation rises monotonically
with rising detuning, cf. Fig.~\ref{fig1}, for large $\Delta$, the degree of
inversion achieved for a finite pulse does no longer increase with $\Delta$,
but eventually decreases again as illustrated in Figs.~\ref{fig2} and \ref{fig3}(a).
To understand this nonmonotonic dependence of the protocol efficiency on $\Delta$,
two things must be born in mind: first, the phonon coupling, that drives the system
towards the stationary state, exhibits a spectral cutoff, i.e., the phonon spectral
density $ J(\omega) = \sum_{\bf{q}} |g_{\bf{q}}|^2 \delta(\omega-\omega_{\bf{q}})$,
that is shown in the inset of Fig.~\ref{fig2}, vanishes for large $\omega$
and second, the strength of the phonon-induced relaxation can be approximately
described by $J(\Omega)$ \cite{machnikowski:04,ramsay:10}. As soon as the
Rabi-frequency $\Omega$ exceeds the frequency $\omega_{\rm{max}}$, where $J(\omega)$
is maximal, the phonon coupling becomes less efficient and the inversion that is
reached for finite pulses is reduced. Further, as $\Omega=\sqrt{f^2+(\Delta/\hbar)^2}$,
the optimal detuning is less than $\hbar\omega_{\rm{max}}$ as it is clearly seen
from Figs.~\ref{fig2} and \ref{fig3}(a).

Obviously, the proposed scheme becomes less efficient when the pulse is too
short and the system is not driven close enough to its stationary state. This
is illustrated in Fig.~\ref{fig3}(b). While pulses longer than those considered
so far do not affect the efficiency, for a pulse length below $10\,\rm{ps}$,
the achieved inversion drops and the protocol is less robust with respect to variations
in the pulse intensity. However, we would like to stress that the pulse length
of $10$ to $15\,\rm{ps}$ that is needed to guarantee a stable phonon-assisted
preparation via off-resonant driving is very much the same as the one that was
needed in recent ARP-based experiments to ensure an adiabatic evolution \cite{wu:11,simon:11}.

\begin{figure}[ttt]
 \includegraphics[width=8.3cm]{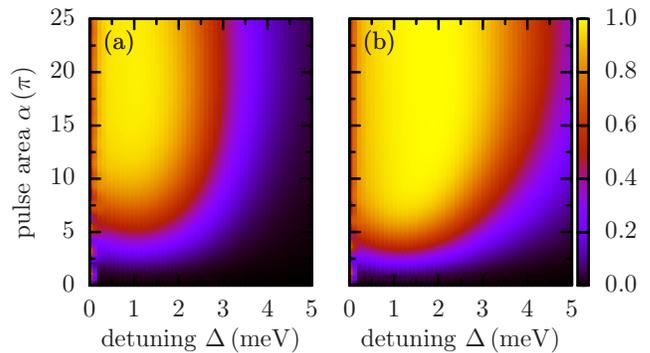}
    \caption{
             (color online) Final exciton occupation $C_X$ after a Gaussian pulse of 15 ps FWHM 
             at $T=4$~K as a function of the detuning $\Delta$ and the pulse area $\alpha$ 
             for (a) GaAs parameters and (b) for a phonon-coupling
             where $|g_{\bf{q}}|^2$ has been increased by a factor of three.
            }
 \label{fig4}
\end{figure}

Unique to the present protocol is that its performance becomes better 
when the strength of the carrier-phonon coupling is increased, thus allowing for 
an almost ideal state preparation in the regime of strong system-environment
interaction, where traditional Rabi flopping or ARP-schemes are known to fail
\cite{foerstner:03,vagov:07,ramsay:10,lueker:12}. Shown in Fig.~\ref{fig4}
is the exciton occupation after a 15 ps lasting pulse (FWHM) as a function
of $\Delta$ and $\alpha$ for the carrier-phonon coupling of GaAs as studied 
so far [Fig.~\ref{fig4}(a)] and for a situation, where we have increased
$|g_{\bf q}|^2$ by a factor of three by hand [Fig.~\ref{fig4}(b)] in order
to roughly simulate the coupling strength of materials like GaN 
\cite{krummheuer:05}. Clearly, the efficiency increases with rising coupling:
the maximal inversion is even closer to one and the range of pulse intensities
and detunings for which high occupations are reached extends considerably.
We stress, that carrying out calculations for strong phonon couplings is a challenge
and that even highly elaborate approximate methods such as a fourth-order correlation
expansion are known to break down in this regime \cite{glaessl:11b}. Here, the
performance of reliable simulations is possible by using a numerically exact
path-integral approach \cite{vagov:11} that accounts fully for all non-Markovian
effects and arbitrary multi-phonon processes.

In light of recent progress in realizing QD-based entangled photon sources
\cite{dousse:10,trotta:12} also the task of preparing the biexciton state
$|XX\rangle$ has become of topical interest. Whether or not the ideas so far
developed for the preparation of the single exciton state can be transferred
to this task is not obvious, because now we have to deal with more electronic
levels and thus a perfect preparation of the biexciton requires that not only 
the ground state but also the single exciton state is completely depopulated.
Furthermore, there is not only a single resonance. In particular, the biexciton
state can be optically excited either by a two-photon process \cite{stufler:06}
that is resonant when $2\hbar\omega=2\hbar\omega_{X}-\Delta_{B}$, where $\Delta_{B}$
is the biexciton binding energy or by a sequential process where first the single
exciton is excited and then a transition from the exciton to the biexciton is
induced. The resonance for the latter process at $\hbar\omega=\hbar\omega_{X}-\Delta_{B}$ 
is known to be a {\em dephasing-induced resonance}, which in $\chi^{(3)}$-signals
shows up only due to the coupling of the electronic system to a bath \cite{axt:96}.

To couple the biexciton we switch the polarization of the exciting pulse from
circular to linear. Then, the light-matter Hamiltonian reads as \footnote{Note,
that here $|X\rangle$ denotes the single exciton state coupled to linearly
polarized light which is different from the exciton state used in Eq.~\eqref{H1}
that couples to circularly polarized light.}
\begin{align}
H_{\rm{QD,\,L}} \!= & \hbar\omega_X |X\rangle\langle X| + (2 \hbar\omega_X \!-\! \Delta_B) |XX\rangle\langle XX| \\ \notag
             &-\!\left[\frac{\hbar f(t)}{2} e^{-i\omega_L t} 
             (|X\rangle\!\langle G|+|XX\rangle\!\langle X|)+\rm{h.c.}\right]\!.
\end{align}
In general $\Delta_B$  can be positive as well as negative but for self-assembled
GaAs QDs takes typically values ranging from $1$ to $3\,\rm{meV}$ \cite{boyle:08,zecherle:10,ramsay:10}.
The carrier-phonon coupling is the same as in Eq.~\eqref{H2} with the only difference
that now the sum over $|\nu\rangle$ runs over three electronic states, $|G\rangle$,
$|X\rangle$, and $|XX\rangle$. 

\begin{figure}[ttt]
 \includegraphics[width=8.5cm]{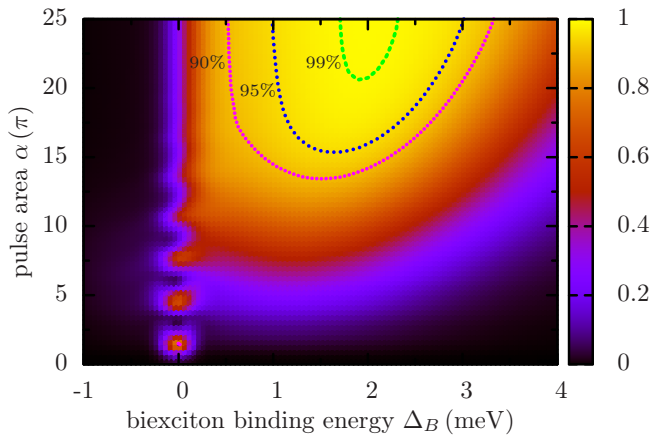}
    \caption{
             (color online) Final biexciton occupation $C_{XX}$ after a Gaussian pulse with a FWHM of 15 ps
             at $T=4$~K as a function of the biexciton binding energy $\Delta_B$ and the pulse
             area $\alpha$. The laser frequency is chosen in resonance with the ground state to
             exciton transition. The contour lines display where certain values of $C_{XX}$ 
             are reached. 
            }
 \label{fig5}
\end{figure}

In the remainder of this Letter we shall demonstrate that an almost perfect
phonon-assisted biexciton preparation is
possible by choosing an excitation that is resonant to the ground-state
to exciton transition provided that the exciton to biexciton transition 
is red-shifted, which typically is the case. This situation is schematically
sketched in the inset of Fig.~\ref{fig1}(b). 

Again, it turns out to be instructive to look first at the
stationary state, that for this excitation is reached due to the acoustic
phonon coupling. Fig.~\ref{fig1}(b) shows the stationary biexciton occupation
as calculated within the path-integral approach for constant driving as a
function of temperature and biexciton binding energy: very high values of
$C_{XX}$ are found for low $T$ and positive $\Delta_B$ indicating that a
phonon-mediated preparation of the biexciton should be possible for the
typical case where the biexciton binding energy shifts the biexciton down
in energy. This supposition is indeed confirmed by the results shown in
Fig.~\ref{fig5}, where the biexciton occupation that is reached after a
Gaussian pulse with $15\,\rm{ps}$ FWHM is plotted as a function of the
pulse area and $\Delta_B$ at $T=4\,\rm{K}$: an almost perfect biexciton
preparation is realized for a wide range of biexciton binding energies
which is robust against variations in the pulse intensity provided that
the pulse is strong enough. It should be noted, that an occupation
probability for $|XX\rangle$ near one implies that due to the action
of the phonons the exciton state as well as the ground state are
practically unoccupied although the ground state to exciton transition
is resonantly driven. As discussed before in detail for the case of the
single exciton preparation, the performance of the scheme becomes worse
when the pulse is chosen too short, but improves with rising strength
of the carrier-phonon coupling (not shown).


In summary, we have presented schemes allowing for an on-demand, high-fidelity,
and robust preparation of the single exciton as well as the biexciton state in 
semiconductor QDs that are based on off-resonant excitations with strong
optical pulses and make active use of the carrier-phonon coupling. 
In particular, we predict that the biexciton can be prepared by
resonantly driving the ground state to exciton transition, implying
that the exciton state will be unoccupied despite its resonant excitation.
The proposed protocols allow for fast state preparations on the time
scale of ~10 ps. We expect our findings to inspire future experimental
research as the regime of off-resonant driving at high intensities is
so far almost unexplored and believe that the proposed phonon-assisted
state preparation schemes can pave the way to more efficient sources
for single or entangled photons. Importantly, the presented protocols
do not only combine the simplicity of Rabi flopping (without the need
of realizing chirped laser pulses) with the robustness of ARP-based schemes,
but perform the better the stronger the carrier phonon coupling is, thus
allowing for an ideal state preparation even in situations with strong
system-environment interaction that are usually thought of as making
control protocols impossible.

M.~G. gratefully acknowledges financial support by the Studienstiftung
des Deutschen Volkes. 


\end{document}